\documentclass[10pt,a4paper,hidelinks, twocolumn]{article}
\usepackage[dvipsnames]{xcolor}
\usepackage{fancyhdr}

\fancypagestyle{CISPA}{

\fancyhead[C]{\includegraphics[width=4cm]{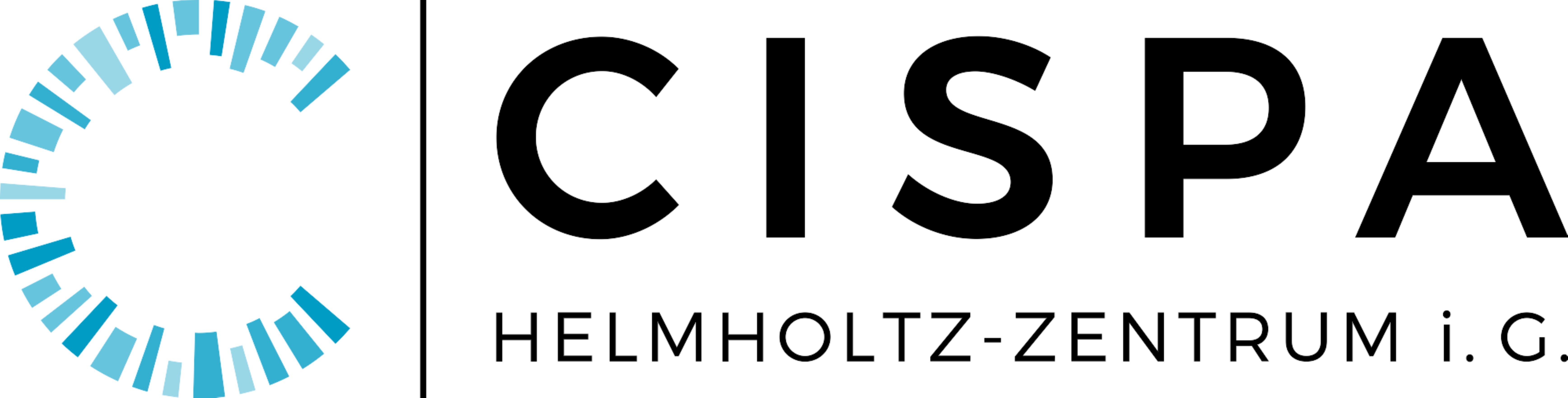}}
}
\usepackage[utf8]{inputenc}
\usepackage{csquotes}
\usepackage[backend=biber, doi=false,isbn=false,url=false, eprint=false]{biblatex}
\usepackage{etoolbox}

\usepackage{eso-pic}
\usepackage[T1]{fontenc}
\usepackage[english]{babel}
\usepackage[nohead,includefoot,margin=2cm]{geometry}
\usepackage[compact,raggedright,small]{titlesec}
\usepackage[affil-it]{authblk} 
\usepackage[runin]{abstract}
\usepackage{titling}
\usepackage{graphicx}
\usepackage{setspace}

\usepackage{booktabs}   %
\usepackage{multirow}
\usepackage{subcaption} %
\usepackage{caption}
\usepackage{tikz}
\usetikzlibrary{arrows}
\usetikzlibrary{positioning}

\usepackage[inline]{enumitem}
\usepackage{microtype} %
\usepackage{alltt}
\usepackage{float}
\usepackage{stfloats}

\usepackage[noend]{algpseudocode}
\usepackage{listings}
\usepackage{relsize} %

\usepackage{utopia}
\usepackage{times}
\usepackage{makecell}
\usepackage{siunitx}
\usepackage{booktabs} %
\usepackage{amsmath}
\usepackage[nameinlink,noabbrev]{cleveref}
\usepackage{multirow}
\usepackage{microtype} %
\usepackage{alltt}
\usepackage{listings}
\usepackage{tikz}
\usepackage{tikz-qtree}     %
\usepackage{balance}

\usepackage{amsthm}
\newtheorem{definition}{Definition}
\usepackage{framed}
\newenvironment{result}{\begin{framed}\centering\it}{\end{framed}}

\usepackage{xspace}
\usepackage%
{cleveref}  %

\usepackage{letltxmacro}
\usepackage{todonotes}
 
\title{ {\LARGE Inputs from Hell} \\[0.5mm] {\LARGE Generating~Uncommon~Inputs~from~Common~Samples}}    %
\date{\small (Dated \today)}
\author%
{Esteban Pavese$^{1}$}
\author%
{Ezekiel Soremekun$^{2}$}
\author%
{Nikolas Havrikov$^{2}$}
\author%
{Lars Grunske$^{1}$}
\author%
{Andreas Zeller$^{2}$}
\affil%
{$^{1}$ Humboldt-Universität zu Berlin, Berlin, Germany\\
\{pavesees, grunske\}@informatik.hu-berlin.de} %
\affil%
{$^{2}$ CISPA / Saarland University, Saarbr\"ucken, Germany\\
\{ezekiel.soremekun, nikolas.havrikov, zeller\}@cispa.saarland}

\LetLtxMacro{\oldtodo}{\todo}
\renewcommand{\todo}[1]{\oldtodo[inline]{#1}}

\def\term#1{\texttt{"\textbf{#1}"}}
\def\nonterm#1{\textnormal{\emph{#1}}}
\def\expandsto{\ensuremath{\rightarrow{}}}
\def\prob#1#2{\textnormal{#1\%}~#2}

\def\|#1|{\textit{#1}}
\def\<#1>{\texttt{#1}}

\def\Gorig{G}           %
\def\Gprob{G_p}         %
\def\Ginvlocal{G_{p^{-1}}}   %

\lstdefinelanguage{JavaScript}{
  keywords={typeof, new, true, false, catch, function, return, null, catch, switch, var, if, in, while, do, else, case, break},
  keywordstyle=\color{blue}\bfseries,
  ndkeywords={class, export, boolean, throw, implements, import, this},
  ndkeywordstyle=\color{darkgray}\bfseries,
  identifierstyle=\color{black},
  sensitive=false,
  comment=[l]{//},
  morecomment=[s]{/*}{*/},
  commentstyle=\color{purple}\ttfamily,
  stringstyle=\color{red}\ttfamily,
  morestring=[b]',
  morestring=[b]"
}

\definecolor{codegreen}{rgb}{0,0.6,0}
\definecolor{codegray}{rgb}{0.5,0.5,0.5}
\definecolor{codepurple}{rgb}{0.58,0,0.82}
\definecolor{backcolour}{rgb}{0.95,0.95,0.92}

\lstdefinestyle{mystyle}{
    commentstyle=\color{codegreen},
    keywordstyle=\color{magenta},
    numberstyle=\tiny\color{codegray},
    stringstyle=\color{codepurple},
    basicstyle=\footnotesize,
    breakatwhitespace=false,
    breaklines=true,
    captionpos=b,
    keepspaces=true,
    numbers=left,
    numbersep=5pt,
    showspaces=false,
    showstringspaces=false,
    showtabs=false,
    tabsize=2
}

\algblockdefx[switch]{Switch}{EndSwitch}%
[1]{\textbf{switch} #1}%
{}%
\algloopdefx{Case}[1]{\textbf{case} #1:}%
\algdef{SE}[DOWHILE]{Do}{doWhile}{\algorithmicdo}[1]{\algorithmicwhile\ #1}%

\def\term#1{\texttt{"\textbf{#1}"}}
\def\nonterm#1{\textnormal{\emph{#1}}}

\def\|#1|{\textit{#1}}
\def\<#1>{\texttt{#1}}

\newcommand{\done}[2]{[#1]\textcolor{blue}{\textbf{DONE}~\textit{#2}}}

\renewcommand{\todo}[1]{}
\renewcommand{\done}[2]{}

\newcommand\BackgroundPic{
    \put(0,0){
    \parbox[b][\paperheight]{\paperwidth}{%
    \vfill
    \centering
    \includegraphics[width=\paperwidth,height=\paperheight]{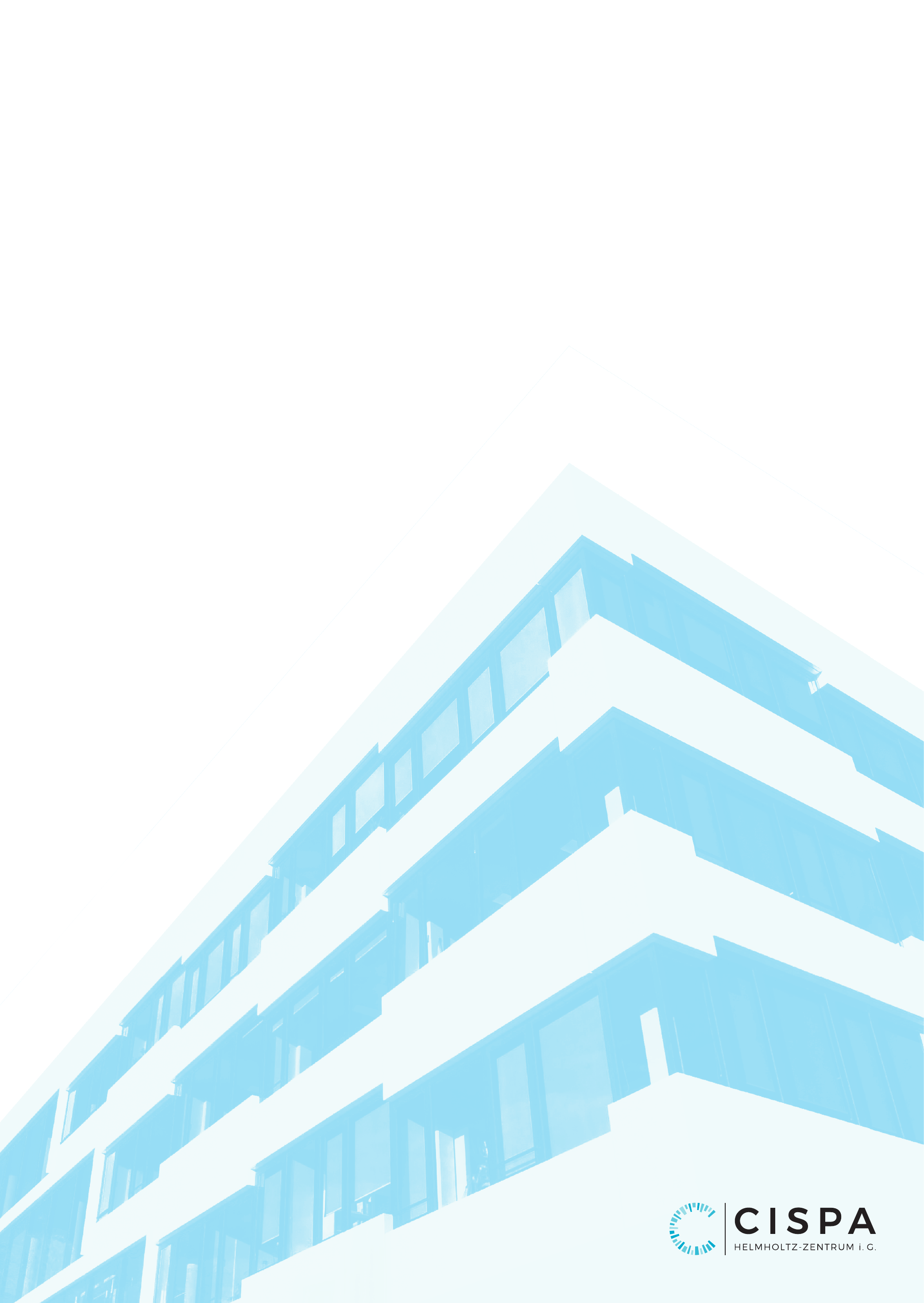}
    \vfill
}}}

\begin{document}
\AddToShipoutPicture*{\BackgroundPic}

\makeatletter
\renewcommand{\Authfont}{\normalsize\sffamily\bfseries}
\renewcommand{\Affilfont}{\normalsize\sffamily\mdseries}

\begin{titlepage}
\newcommand{\HRule}{\rule{\linewidth}{0.1mm}}
\centering
  \textsc{\LARGE {\fontfamily{Montserrat-TOsF}\selectfont CISPA Helmholtz-Zentrum i.G.}}\\[1.5cm]

  \vspace{2.4 cm}
  \HRule \\[0.2cm]
  {\huge\sffamily\bfseries \@title\par}
  \vspace{0.2cm}
  \HRule \\[1.5cm]

  {\sffamily \@author\par}
\vfill

\end{titlepage}

\twocolumn[
\makeatother
\setlength{\affilsep}{0.1em}
\addto{\Affilfont}{\small}
\renewcommand{\Authfont}{\normalsize}
\renewcommand{\Affilfont}{\normalsize}

\pretitle{\begin{center}\large\bfseries}
\posttitle{\end{center}}
\vspace{0.1in}
\maketitle
\thispagestyle{CISPA}

\begin{abstract}
Generating structured input files to test programs can be performed by techniques that produce them from a grammar that serves as the specification for syntactically correct input files.
Two interesting scenarios then arise for effective testing.
In the first scenario, software engineers would like to generate inputs that are as \emph{similar} as possible to the inputs in common usage of the program, to test the reliability of the program.
More interesting is the second scenario where inputs should be as \emph{dissimilar} as possible from normal usage.
This is useful for robustness testing and exploring yet uncovered behavior.
To provide test cases for both scenarios, we leverage a context-free grammar to parse a set of sample input files that represent the program's common usage, and determine \emph{probabilities} for individual grammar production as they occur during parsing the inputs.
\emph{Replicating} these probabilities during grammar-based test input generation, we obtain inputs that are close to the samples.
\emph{Inverting} these probabilities yields inputs that are strongly dissimilar to common inputs, yet still valid with respect to the grammar. %
Our evaluation on three common input formats (JSON, JavaScript, CSS) shows the effectiveness of these approaches in obtaining instances from both sets of inputs.
\end{abstract}
]

\section{Introduction}
\label{sec:intro}
During the process of software testing, software engineers typically look at satisfying two goals.
First, ensuring that the software works well on \emph{common} inputs, such that the software delivers its promise on the vast majority of cases (and for the vast majority of customers) that will be seen in typical operation.
This is usually achieved by having a set of tests (manually written or generated) that covers this common behavior.
Besides these \emph{common} inputs, though, it is also advisable to test for \emph{uncommon} inputs.
The rationale for this is that such inputs would exercise code that is less frequently used in production, possibly less tested, and possibly less well understood~\cite{cadar2008GPDE}.

The question that then arises is, how can engineers obtain such uncommon inputs?
In this paper, we focus on the problem of generating uncommon (but otherwise syntactically correct and perfectly legal) inputs that are \emph{unlikely to be seen in typical operation}.
To this end, we assume the existence of a \emph{context-free grammar} that describes the input language to a program, that is, it describes the set of its valid inputs.
Using such a grammar, we can \emph{parse} existing common input samples and \emph{count} how frequently specific elements occur in these samples.
Armed with these numbers, we can enrich the grammar to become a \emph{probabilistic grammar}, in which choices present in productions carry different likelihoods.
Since these probabilities come from the common samples used for the quantification, this grammar describes the distribution of valid, but \emph{common} inputs.
The key idea is that now we can \emph{invert} these probabilities in order to obtain a second probabilistic grammar.
This \emph{inverted} grammar, however, describes in turn the distribution of legal, but \emph{uncommon} inputs. We call them ``inputs from hell''.

As an example of such ``inputs from hell'', consider \Cref{fig:hell_rhino}, listing two JavaScript inputs generated by focusing on uncommon features.
Both of these snippets are valid JavaScript code, but cause the Mozilla Rhino JavaScript compiler to crash during interpretation.
They make use of so-called \emph{destructuring assignments}: in JavaScript, it is allowed to have several variables on the left hand side of an assignment or initialization.
In such a case, each gets assigned a part of the structure on the right hand side, as in
\begin{lstlisting}[language=JavaScript]
    var [one, two, three] = [1, 2, 3];
\end{lstlisting}
where the variable \<one> is assigned a value of \<1>, \<two> a value of \<2>, and so on. Such destructuring assignments, although useful in some contexts, are not extensively found across JavaScript samples and tests.This is precisely why the aim of our approach is to generate these ``inputs from hell''.

\begin{figure}[hb]
\begin{lstlisting}[language=JavaScript]
    const [c, y, y] = [];
    var { a: {} = 'b' } = {};
\end{lstlisting}
\caption{Two inputs from hell that break Rhino~1.7.7.2}
\label{fig:hell_rhino}
\end{figure}

This paper makes the following contributions:

\begin{enumerate}
\item We show how to use context-free grammars to determine production probabilities from a given set of input samples.%

\item We show how to use mined probabilities to produce inputs that are \emph{similar to a set of given samples}.
This is useful for thoroughly testing commonly used features (regression testing), or to test the surroundings of previously failure-inducing inputs.
As a result our approach leverages the well-known concept of probabilistic grammars for both mining and test case generation.
In our evaluation using the JSON, CSS and JavaScript formats, we show that our approach repeatedly covers the same code as the original sample inputs.

\item We show how to use mined probabilities to produce inputs that are \emph{markedly dissimilar} to a set of given samples, yet still valid according to the grammar.
This is useful for robustness testing, as well as for exploring program behavior not triggered by the sample inputs.
We are not aware of any other technique that achieves this objective.
In our evaluation using the same subjects, we show that our approach is successful in repeatedly covering code not covered in the original samples.
\end{enumerate}

\section{Inputs from Hell in a Nutshell}
\label{sec:motivation}

To demonstrate how we produce both common and uncommon inputs, let us illustrate our approach using a simple example grammar.
Let us assume we have a program~$P$ that processes \emph{arithmetic expressions}; its inputs follow the standard syntax given by the grammar~$\Gorig$ below.
{\small
\begin{alltt}
	\nonterm{Expr} \expandsto \nonterm{Term} | \nonterm{Expr} \term{+} \nonterm{Term} | \nonterm{Expr} \term{-} \nonterm{Term};
	\nonterm{Term} \expandsto \nonterm{Factor} | \nonterm{Term} \term{*} \nonterm{Factor} 
        | \nonterm{Term} \term{/} \nonterm{Factor};
	\nonterm{Factor} \expandsto \nonterm{Int} | \term{+} \nonterm{Factor}
	    | \term{-} \nonterm{Factor} | \term{(} \nonterm{Expr} \term{)};
	\nonterm{Int} \expandsto \nonterm{Digit} \nonterm{Int} | \nonterm{Digit};
	\nonterm{Digit} \expandsto \term{0} | \term{1} | \term{2} | \term{3} | \dots | \term{9};
\end{alltt}
}
Let us further assume we have discovered a bug in~$P$: the input $I = \<1 * (2 + 3)>$ is not evaluated properly.
We have fixed the bug in~$P$, but want to ensure that similar inputs would also be handled in a proper manner.

To obtain inputs \emph{similar} to~$I$, we first use the grammar~$\Gorig$ to \emph{parse} $I$ and determine the \emph{distribution} of the individual choices in productions.
This makes~$\Gorig$ a \emph{probabilistic} grammar~$\Gprob$ in which the productions' choices are tagged with their probabilities.
For the input $I$ above, for instance, we obtain the probabilistic rule

{\small
\begin{alltt}
\nonterm{Digit} \expandsto \prob{0}{\term{0}} | \prob{33.3}{\term{1}} | \prob{33.3}{\term{2}}
    | \prob{33.3}{\term{3}} | \prob{0}{\term{4}} | \prob{0}{\term{5}}
    | \prob{0}{\term{6}} | \prob{0}{\term{7}} | \prob{0}{\term{8}} | \prob{0}{\term{9}};
\end{alltt}
}%

which indicates the distribution of digits in~$I$.
Using this rule for production, we would obtain ones, twos, and threes at equal probabilities, but none of the other digits.
\Cref{fig:prob-grammar} shows the grammar~$\Gprob$ as extension of~$\Gorig$ with all probabilities as extracted from the derivation tree of~$I$ (\Cref{fig:expr-derivation}).
In this derivation tree we see, for instance, that the nonterminal \nonterm{Factor} occurs 4 times in total.
75\% of the time it produces integers (\nonterm{Int}), while in the remaining 25\% it produces a parenthesis expression (\term{(} \nonterm{Expr} \term{)}).
Expressions using unary operators like \term{+} \nonterm{Factor} and \term{-} \nonterm{Factor} do not occur.

\begin{figure}[!h]
	{\small
		\begin{tikzpicture}
		\tikzset{level distance=16pt}
		\Tree [.\nonterm{Expr}
		[.\nonterm{Expr}
		[.\nonterm{Term}
		[.\nonterm{Factor}
		[.\nonterm{Int}
		[.\nonterm{Digit}
		\term{1}
		]
		]
		]
		]
		]
		\term{+}
		[.\nonterm{Term}
		[.\nonterm{Factor}
		\term{(}
		[.\nonterm{Expr}
		[.\nonterm{Term}
		[.\nonterm{Term}
		[.\nonterm{Factor}
		[.\nonterm{Int}
		[.\nonterm{Digit}
		\term{2}
		]
		]
		]
		]
		\term{*}
		[.\nonterm{Factor}
		[.\nonterm{Int}
		[.\nonterm{Digit}
		\term{3}
		]
		]
		]
		]
		]
		\term{)}
		]
		]
		]
		\end{tikzpicture}
	}%
	\caption{Derivation tree representing \term{1 + (2 * 3)}}
	\label{fig:expr-derivation}
\end{figure}

If we use~$\Gprob$ from \Cref{fig:prob-grammar} as a probabilistic production grammar, we obtain inputs according to these probabilities.
As listed in \Cref{fig:prob-samples}, these inputs uniquely consist of the digits and operators seen in our sample \<1 * (2 + 3)>.
All of these inputs are likely to cover the same code in~$P$ as the original sample input, yet with different input structures that trigger the same functionality in~$P$ in several new ways.

\begin{figure}[!h]
	{\small
		\begin{alltt}
			\nonterm{Expr} \expandsto \prob{66.7}{\nonterm{Term}} | \prob{33.3}{\nonterm{Expr} \term{+} \nonterm{Term}}
			    | \prob{0}{\nonterm{Expr} \term{-} \nonterm{Term}};
			\nonterm{Term} \expandsto \prob{75}{\nonterm{Factor}} | \prob{25}{\nonterm{Term} \term{*} \nonterm{Factor}}
			    | \prob{0}{\nonterm{Term} \term{/} \nonterm{Factor}};
			\nonterm{Factor} \expandsto \prob{75}{\nonterm{Int}} | \prob{0}{\term{+} \nonterm{Factor}}
			    | \prob{0}{\term{-} \nonterm{Factor}} | \prob{25}{\term{(} \nonterm{Expr} \term{)}};
			\nonterm{Int} \expandsto \prob{0}{\nonterm{Digit} \nonterm{Int}} | \prob{100}{\nonterm{Digit}};
			\nonterm{Digit} \expandsto \prob{0}{\term{0}} | \prob{33.3}{\term{1}} | \prob{33.3}{\term{2}}
			    | \prob{33.3}{\term{3}} | \prob{0}{\term{4}} | \prob{0}{\term{5}}
			    | \prob{0}{\term{6}} | \prob{0}{\term{7}} | \prob{0}{\term{8}} | \prob{0}{\term{9}};
		\end{alltt}
	}\vspace{-0.2cm}
	\caption{Probabilistic grammar~$\Gprob$, expanding $\Gorig$}
	\label{fig:prob-grammar}
\end{figure}

\begin{figure}[htb]
	{\small
		\begin{alltt}
			(2 * 3)
			2 + 2 + 1 * (1) + 2
			((3 * 3))
			3 * (((3 + 3 + 3) * (2 * 3 + 3))) * (3)
			3 * 3
			3 * 1 * 3
			((3) + 2 + 2 * 1) * (1)
			1
			((2)) + 3
		\end{alltt}
	}%
	\caption{Inputs generated from~$\Gprob$ in \Cref{fig:prob-grammar}}
	\label{fig:prob-samples}
\end{figure}

Replicating ``more of the same'' features as found in sample inputs makes most sense if these studied inputs can be associated with errors.
However if we, as in most cases, only have sample inputs that work just fine, we would typically be interested in inputs that are \emph{different} from our samples.
We can easily obtain such inputs by \emph{inverting} the mined probabilities: if a rule previously had a weight of~$p$, we now assign it a weight of~$1 / p$, normalized across all production alternatives.
For our \emph{Digit} rule, this gives the digits not seen so far a weight of $1/0 = \infty$, which is still distributed equally across all seven alternatives, yielding individual probabilities of %
$1/7 = 14.3\%$.
Proportionally, the weights for the digits already seen in~$I$ are infinitely small, yielding a probability of effectively zero.  The thus ``inverted'' rule reads now:

{\small
\begin{alltt}
\nonterm{Digit} \expandsto \prob{14.3}{\term{0}} | \prob{0}{\term{1}} | \prob{0}{\term{2}} | \prob{0}{\term{3}}
    | \prob{14.3}{\term{4}} | \prob{14.3}{\term{5}} | \prob{14.3}{\term{6}}
    | \prob{14.3}{\term{7}} | \prob{14.3}{\term{8}} | \prob{14.3}{\term{9}};

\end{alltt}
}%

Applying this inversion to rules with non-terminal symbols is equally straightforward.
The resulting probabilistic grammar~$\Ginvlocal$ is given in \Cref{fig:inv-local-prob-grammar}.

\begin{figure}[htb]
	{\small
		\begin{alltt}
			\nonterm{Expr} \expandsto \prob{0}{\nonterm{Term}} | \prob{0}{\nonterm{Expr} \term{+} \nonterm{Term}}
			    | \prob{100}{\nonterm{Expr} \term{-} \nonterm{Term}};
			\nonterm{Term} \expandsto \prob{0}{\nonterm{Factor}} | \prob{0}{\nonterm{Term} \term{*} \nonterm{Factor}}
			    | \prob{100}{\nonterm{Term} \term{/} \nonterm{Factor}};
			\nonterm{Factor} \expandsto \prob{0}{\nonterm{Int}} | \prob{50}{\term{+} \nonterm{Factor}}
			    | \prob{50}{\term{-} \nonterm{Factor}} | \prob{0}{\term{(} \nonterm{Expr} \term{)}};
			\nonterm{Int} \expandsto \prob{100}{\nonterm{Digit} \nonterm{Int}} | \prob{0}{\nonterm{Digit}};
			\nonterm{Digit} \expandsto \prob{14.3}{\term{0}} | \prob{0}{\term{1}} | \prob{0}{\term{2}} | \prob{0}{\term{3}}
			    | \prob{14.3}{\term{4}} | \prob{14.3}{\term{5}} | \prob{14.3}{\term{6}}
			    | \prob{14.3}{\term{7}} | \prob{14.3}{\term{8}} | \prob{14.3}{\term{9}};
		\end{alltt}
	}%
	\caption{Grammar~$\Ginvlocal$ inverted from~$\Gprob$ in \Cref{fig:prob-grammar}}
	\label{fig:inv-local-prob-grammar}
\end{figure}

This inversion can lead to infinite derivations, for example, the production rule in $\Ginvlocal$ for generating \nonterm{Expr} is recursive 100\% of the time, expanding only to \nonterm{Expr} \term{-} \nonterm{Term}, without chance of hitting the base case.
As a result, we take special measures to avoid such infinite productions during input generation, which we will detail further on.

If we use~$\Ginvlocal$ as a production grammar---and avoiding infinite production---we obtain inputs as shown in \Cref{fig:inv-local-prob-samples}.
These inputs now focus on operators like subtraction or division or unary operators not seen in our input samples.
Likewise, the newly generated digits cover the complement of those digits previously seen.
Yet, all inputs are syntactically valid according to the grammar.
With both the sets of similar and dissimilar inputs, we can expect to have a good set of regression tests as well as a set exploiting less frequently used functionality.

\begin{figure}[htb]
	{\small
		\begin{alltt}
			+5 / -5 / 7 - +0 / 6 / 6 - 6 / 8 - 5 - 4
			-4 / +7 / 5 - 4 / 7 / 4 - 6 / 0 - 5 - 0
			+5 / ++4 / 4 - 8 / 8 - 4 / 8 / 7 - 8 - 9
			-6 / 9 / 5 / 8 - +7 / -9 / 6 - 4 - 4 - 6
			+8 / ++8 / 5 / 4 / 0 - 5 - 4 / 8 - 8 - 8
			-9 / -5 / 9 / 4 - -9 / 0 / 5 - 8 / 4 - 6
			++7 / 9 / 5 - +8 / +9 / 7 / 7 - 6 - 8 - 4
			-+6 / -8 / 9 / 6 - 5 / 0 - 5 - 8 - 0 - 5
		\end{alltt}
	}\vspace{-0.2cm}
	\caption{Inputs generated from $\Ginvlocal$ from \Cref{fig:inv-local-prob-grammar}}
	\label{fig:inv-local-prob-samples}
\end{figure}

\section{Approach}
\label{sec:approach}

In order to explain our approach in detail, we start with introducing basic notions of probabilistic grammars.

\subsection{Probabilistic Grammars}

The probabilistic grammars that we employ in this paper are based on the well-known context-free grammars (CFGs)~\cite{Ullman01}.
\begin{definition}[Context-free grammar]
	A context-free grammar is a 4-tuple $(V,T,P,S_0)$, where $V$ is the set of \emph{non-terminal symbols}, $T$ the terminals, $P: V \rightarrow (V \cup T)^{*}$ the set of productions, and $S_0 \in V$ the start symbol.
\end{definition}

In a \emph{non-probabilistic grammar}, rules for a non-terminal symbol~$S$ provide $n$~alternatives $A_i$ for expansion:
\begin{equation}
S \expandsto A_1\:\:|\:\:A_2\:\:|\: \dots \:|\:\: A_n
\label{eq:nonprob-grammar}
\end{equation}

In a \emph{probabilistic grammar}, each of the alternatives~$A_i$ in~\Cref{eq:nonprob-grammar} is augmented with a probability $p_i$, where $\sum_{i=1}^{n}p_i = 1$ holds:
\begin{equation}
S \expandsto p_1\:A_1 \:\:|\:\: p_2\:A_2 \:\:|\: \dots \:|\:\: p_n\:A_n
\label{eq:prob-grammar}
\end{equation}

If we are using these grammars for generation of a sentence of the language described by the grammar, each alternative~$A_i$ has a probability of~$p_i$ to be selected when expanding~$S$.

By convention, if one or more $p_i$ are not specified in a rule, we assume that their value is the complement probability, distributed equally over all alternatives with these unspecified probabilities.
Consider the rule
\begin{alltt}
\nonterm{Letter} \expandsto \prob{40.0}{\term{a}} | \term{b} | \term{c}
\end{alltt}
Here, the probabilities for \term{b} and \term{c} are not specified; we assume that the complement from \term{a}, namely 60\%, is equally distributed over them, yielding effectively
\begin{alltt}
\nonterm{Letter} \expandsto \prob{40.0}{\term{a}} | \prob{30.0}{\term{b}} | \prob{30.0}{\term{c}}
\end{alltt}
Formally, to assign a probability to an unspecified~$p_i$, we use
\begin{equation}
p_i = \frac{1 - \sum\{p_j | \text{$p_j$ is specified for $A_j$}\}}%
{\text{number of alternatives $A_k$ with unspecified~$p_k$}}
\label{eq:unspecified-p}
\end{equation}
Again, this causes the invariant $\sum_{i=1}^{n}p_i = 1$ to hold.  If no $p_i$ is specified for a rule with $n$ alternatives, as in \Cref{eq:nonprob-grammar}, then \Cref{eq:unspecified-p} makes each $p_i = 1 / n$, as intended.

\subsection{Learning Probabilities}

Our aim now is to turn a classical context-free grammar~$\Gorig$ into a probabilistic grammar~$\Gprob$ capturing the probabilities from a set of samples.
That is, to determine the necessary $p_i$ values as defined in \Cref{eq:prob-grammar} from these samples.
This is achieved by \emph{counting} how frequently individual alternatives occur during parsing in each production context, and then to determine appropriate probabilities.

In language theory, the result of parsing a sample input~$I$ using~$\Gorig$ is a \emph{derivation tree}~\cite{trees}, representing the structure of a sentence according to~$\Gorig$.
As an example, consider \Cref{fig:expr-derivation}, representing the input \term{1 + (2 * 3)} according to the example arithmetic expression grammar in \Cref{sec:motivation}.
In this derivation tree, we can now \emph{count} how frequently a particular alternative $A_i$ was chosen in the grammar~$\Gorig$ during parsing.
In \Cref{fig:expr-derivation}, the rule for \nonterm{Expr} is invoked three times during parsing.
This rule expands once (33.3\%) into \nonterm{Expr} \term{+} \nonterm{Term} (at the root); and twice (66.7\%) into \nonterm{Term} in the subtrees.
Likewise, the \nonterm{Term} symbol expands once (25\%) into \nonterm{Term} \term{*} \nonterm{Factor} and three times (75\%) into \nonterm{Factor}.

Formally, given a set $T$ of derivation trees from a grammar~$\Gorig$ applied on sample inputs, we determine the probabilities $p_i$ for each alternative $A_i$ of a symbol~$S \expandsto A_1\:\:| \dots |\:\:A_n$ as
\begin{equation}
p_i = \frac{\text{Expansions of $S \expandsto A_i$ in $T$}}{\text{Expansions of $S$ in $T$}}
\label{eq:learning-p}
\end{equation}
If a symbol~$S$ does not occur in~$T$, then \Cref{eq:learning-p} makes $p_i = 0/0$ for all alternatives~$A_i$; in this case, we treat all~$p_i$ for~$S$ as \emph{unspecified,} assigning them a value of $p_i = 1/n$ in line with \Cref{eq:unspecified-p}.

In our example, \Cref{eq:learning-p} yields the probabilistic grammar~$\Gprob$ in \Cref{fig:prob-grammar}, assigning probabilities to all alternatives.

\subsection{Inverting Probabilities}

We turn our attention now to the converse approach; namely producing inputs that \emph{deviate} from the sample inputs that were used to learn the probabilities described above.
This ``less of the same'' approach promises to be useful if we accept that our samples are not able to cover all the possible behavior of the system under test, and if we want to find bugs in behaviors that are either not exercised by our samples, or do so rarely.

The key idea is to \emph{invert} the probability distributions as learned from the samples, such that the input generation focuses on the complement section of the language (w.r.t. the samples and those inputs generated by the probabilistic grammar).
If some symbol occurs frequently in the parse trees corresponding to the samples, this approach should generate the symbol less frequently, and vice versa: if the symbol seldom occurs, then the approach should definitely generate it often.

For a moment, let us ignore probabilities and focus on \emph{weights} instead.
That is, the absolute (rather than relative) number of occurrences of a symbol in the parse tree of a sample.
We start by determining the occurrences of a symbol~$A$ during a production $S$ found in a derivation tree~$T$:

\begin{equation}
w_{A,S} =  \begin{array}{l}
\text{Occurrence count of $A$ in the }\\
\text{expansions of symbol $S$ in $T$}
\end{array}
\label{eq:learning-w}
\end{equation}

To obtain \emph{inverted} weights $w'_{A,S}$, a simple way is to make each~$w'_{A,S}$ based on the reciprocal value of~$w_{A,S}$, that is

\begin{equation}
w'_{A,S} = {w_{A,S}}^{-1} = \frac{1}{w_{A,S}}
\label{eq:inversion}
\end{equation}

If the set of samples is small enough, or focuses only on a section of the language of the grammar, it might be the case that some production or symbol never appears in the parsing trees.
If this is the case, then the previous equations end up yielding $w_{A,S} = 0$.
We can compute ${w_{A,S}}^{-1} = \infty$, assigning the elements not seen an infinite weight.
Consequently, all symbols~$B$ that were indeed seen before (with $w_{B,S} > 0$) are assigned an infinitesimally small weight, leading to $w_{B,S}' = 0$.
The remaining infinite weight is then distributed over all of the originally ``unseen'' elements with original weight $w_{A,S} = 0$.
Recall the arithmetic expression grammar in \Cref{sec:motivation}; such a situation arises when we consider the rule for the symbol \nonterm{Digit}: the inverted probabilities for the rule focus exclusively on the complement of the digits seen in the sample.

All that remains in order to obtain actual probabilities is to \emph{normalize} the weights back into a probability measure, ensuring for each production rule that its invariant $\sum_{i=1}^{n}p_i' = 1$ holds:

\begin{equation}
p'_i = \frac{w_i'}{\sum_{i=1}^{n}w_i'}
\label{eq:inverted}
\end{equation}

\subsection{Producing Inputs from a Grammar}

Given a probabilistic grammar $\Gprob$ for some language (irrespective of whether it was obtained by learning from samples, by inverting, or simply written that way in the first place), our next step in the approach is to generate inputs following the specified productions.
This generation process is actually very simple, since it reduces to produce instances by traversing the grammar, as if it were a Markov chain.
However, this generation runs the serious risk of probabilistically choosing productions that lead to an excessively large parsing tree.
Even worse, the risk of generating an \emph{unbounded} tree is very real, as can be seen in the rule for the symbol \nonterm{Int} in the arithmetic expression grammar in \Cref{sec:motivation}.
The production rule for said symbol triggers, with probability 1.0, a recursion with no base case, and will never terminate.

Our inspiration for constraining the growth of the tree during input generation comes from the PTC2 algorithm~\cite{PTC2}.
The main idea of this algorithm is to allow the expansion of not-yet-expanded productions, but all the while ensuring that the number of productions does not exceed a certain threshold of performed expansions.
This threshold would be set as parameter of the input generation process.
Once this threshold is exceeded, the partially generated instance cannot be truncated, as that would result in an illegal input.
Alternatively, we choose to allow further expansion of the necessary non-terminal symbols.
However, from this point on, expansions are not chosen probabilistically.
Rather, the choice is constrained to those expansions that generate the \emph{shortest} possible expansion tree.
This ensures both termination of the generation procedure, as well as trying to keep the input size close to the threshold parameter.
This choice, however, does introduce a bias that may constitute a threat to the validity of our experiments.
We discuss this issue later in \Cref{sec:validation}.

\subsection{Implementation}

\begin{figure*}[ht]
	\centering
	\includegraphics[width=\textwidth]{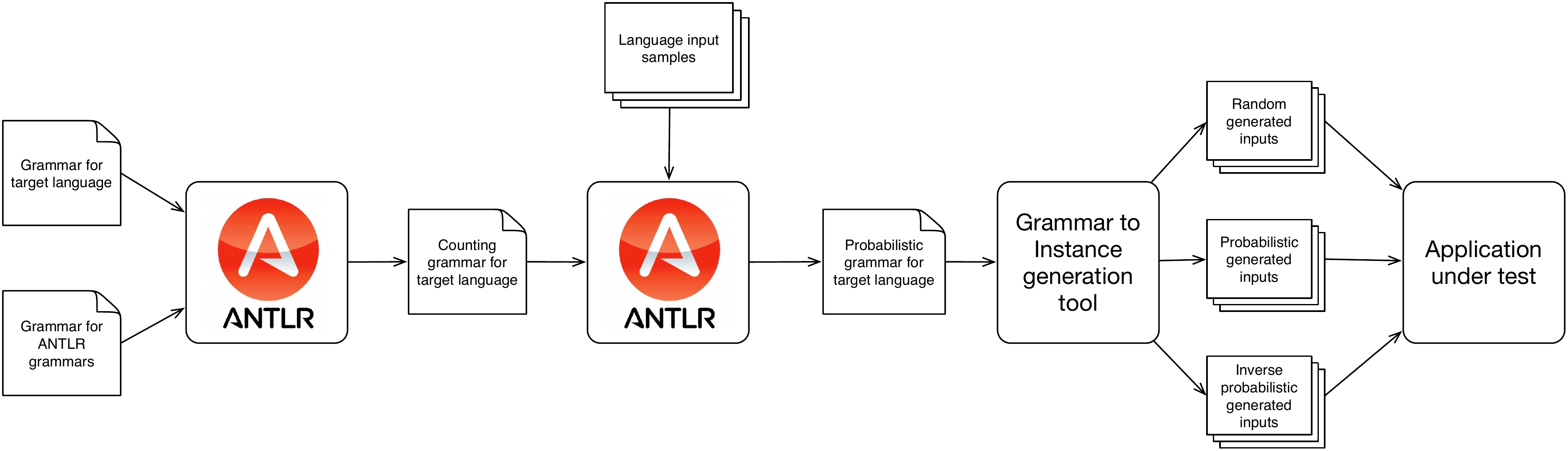}
        \vspace{-\baselineskip}
	\caption{Workflow for the generation of \emph{more of the same} and \emph{less of the same}.}
	\label{fig:approach_implementation}
\end{figure*}

As a prerequisite for carrying out our approach, we only assume we have the context-free grammar of the language available for which we are interested in generating inputs, and a collection (no matter the size) of inputs that we will assume are \emph{common} inputs.
Armed with these elements, we perform the workflow detailed in \Cref{fig:approach_implementation}.

The first step of the approach is to obtain a \emph{counting grammar} from the original grammar.
This counting grammar is, from the parsing point of view, completely equivalent to the original grammar.
However, it is augmented with \emph{actions} during parsing which perform all necessary counting of symbol occurrences parallel to the parsing phase.
Finally, it outputs the probabilistic grammar.
Note that this first phase requires not only the grammar of the target language, but also the grammar of the \emph{language in which the grammar itself is written}.
That is, generating the probabilistic grammar not only requires parsing sample inputs, but also the grammar itself.
In the particular case of our implementation, we make use of the well-known parser generator ANTLR~\cite{ANTLR}.

Once the probabilistic grammar is obtained, we derive the probabilistically-inverted grammar as described in this section.
Armed with both probabilistically annotated grammars, we can continue with the input generation procedure.

\section{Experimental Evaluation}
\label{sec:validation}

In this section we evaluate our approach by applying the technique to several case studies.
In particular, we ask the following research questions:
\begin{itemize}
	\item[] \textbf{(RQ1)} Can a learned grammar be used to generate inputs that resemble those that were employed during the grammar training? (``more of the same'')
	\item[] \textbf{(RQ2)} Can a learned grammar be modified so it can generate inputs that, opposed to \textbf{(RQ1)}, are \emph{in contrast} to those employed during the grammar training? (``less of the same'')
\end{itemize}

To answer the first two questions, we need to compare inputs in order to decide whether these inputs are ``similar'' or ``contrasting''.
In the scope of this evaluation, we will use the \emph{method call frequency} as a measure of input similarity.
We will define this measure later in this section, and we will discuss its usefulness, as well as alternatives to it, when we discuss the threats to the validity of our validation approach.

\subsection{Evaluation Setup}
\subsubsection{Generated Inputs}
Once a probabilistic grammar is learned from the training instances, we generate several inputs that are fed to each subject. Our evaluation involves the generation of two types of test suites:
\begin{enumerate}[label={\alph*)}]
	\item \emph{Probabilistic} - choice between productions is governed by the distribution specified by the learned probabilities in the grammar.
	\item \emph{Inverse} - choice is governed by the distribution obtained as a result of the inversion process described in Section~\ref{sec:approach}.
\end{enumerate}
Expansion size control is carried out in order to avoid unbounded expansion as described in \Cref{sec:approach}.

\subsubsection{Research Protocol}
In our evaluation, we generate test suites and measure the frequency of method calls they induce in our subject programs.
We use the HPROF~\cite{HPROF} profiler to accurately monitor the number of method calls for each input, since all subjects are implemented in Java.
For each input language, the experimental protocol proceeds as follows:
\begin{enumerate}[label={\alph*)}]
\item We randomly select five files from a pool of thousands of sample files crawled from GitHub code repositories, and through our approach produce a probabilistic grammar out of them.
\item We feed the sampled input files into the subject program and record the frequency of method calls using HPROF~\cite{HPROF}.
\item Using the probabilistic grammar, we generate test suites, each one containing 100 input files.%
We generate a total of 1000 test suites, in order to control for variance in the input files.
Overall, each experiment contains 100,000 input files (100 files x 1,000 runs).
We perform this step for both probabilistic and inverse generations.
Hence, the total number of inputs generated for each grammar is 200,000 (1,000 suites of 100 inputs each, a set of suites for each experiment). %
\item We test each subject program, by feeding the input files into the subject program and recording the frequency of method calls using HPROF~\cite{HPROF}.
\end{enumerate}
All experiments were conducted on a shared server with 64 cores and 126 GB of RAM; more specifically an Intel(R) Xeon(R) CPU E5-2683 v4 @ 2.10GHz with 64 virtual cores (Intel Hyperthreading), running Debian 9.5 Linux.

\subsubsection{Subject Programs}
In order to validate our approach, we evaluated the technique by generating inputs and feeding them to a variety of Java applications.
All these applications are open source programs using three different input formats, namely JSON, JavaScript and CSS3.
\Cref{tbl:validation_subjects} summarizes the subjects to be analyzed, their input format and the number of methods in each implementation.

\begin{table}[ht]
\centering
\begin{tabular}{|c|c|c|}
	\hline
	\textbf{Subject} & \textbf{\# of methods} & \textbf{Input language} \\
	\hline
	Argo & 523 & \multirow{9}{*}{JSON} \\
	Genson & 1182 & \\
	Gson & 793 & \\
	JSONJava & 202 & \\
	Jackson & 5378 &  \\
	JsonToJava & 294 & \\
	MinimalJson & 224 & \\
	Pojo & 445 & \\
	json-simple & 63 & \\
	\hline
	Rhino & 4873  & JavaScript \\
	\hline
	cssValidator & 7774 & CSS3 \\
	\hline
\end{tabular}
\caption{Subjects and input language for approach validation.}
\label{tbl:validation_subjects}
\end{table}

The initial, unquantified grammars for the input subjects were adapted from those in the repository of the well-known parser generator ANTLR~\cite{ANTLR}\footnote{The original grammars can be found at \url{https://github.com/antlr/grammars-v4}. }.
Training samples were obtained by scraping GitHub repositories for the required format files.
The probabilistic grammars developed from the original ones, as well as the obtained training samples can be found in the artefact package submitted along this paper.

\subsubsection{Measuring (dis)similarity}

Questions \textbf{(RQ1)} and \textbf{(RQ2)} refer to a notion of similarity between inputs.
Although white-box approaches exist that aim to measure test-case (dis)similarity~\cite{FeldtTGA08,ShiCFFX16}, applying them to complex grammar-based inputs is not straightforward.
However, in this paper, since we are dealing with evaluating the behavior of a certain piece of software, it makes sense to aim for a notion of \emph{semantic} similarity.
In this sense, two inputs are semantically similar if they incite similar behaviors in the software that processes them.
In order to achieve this, we define a measure of input similarity in terms of their method call frequency.
We will say two inputs are similar if they trigger a similar distribution in the frequency with which the methods of the piece of software under analysis are called.
Of course, such a notion allows for a great variance drift if we were to compare only two inputs.
Therefore, we perform this comparison on test suites as a whole to dampen the effect of this variance.

Using this proxy measure of method call frequency, we will aim at answering \textbf{(RQ1)} and \textbf{(RQ2)}.
The first question will be answered satisfactorily if the distribution of call frequencies when running the subjects on \emph{probabilistically} generated suites is similar to the frequency when running the software on the \emph{training} samples.
Likewise, the second question will be answered positively if the call frequency distributions for suites generated with the \emph{inverse} approach are markedly different.

\subsection{Experimental results}

In the figures below ranging from \Cref{fig:json_results_gson} to \Cref{fig:css_results_cssvalidator}, we show a representative selection of our results\footnote{The full range of charts is omitted for space reasons. However, all charts, as well as the raw data, are available as part of the artifact package. Moreover, the charts shown here have been selected so that they are representative of the whole set; that is, the omitted charts do not deviate significantly.}.
In this section, we describe the data depicted in these charts and offer our interpretations.

For each subject, two charts are constructed.
In both charts, the horizontal axis (which is otherwise unlabelled) represents the set of methods in the subject, ordered by the \emph{frequency of calls} in the experiment on \emph{probabilistic inputs}.
The chart at the top represents the \emph{accumulated} call frequency for each strategy (calls in the samples, in probabilistic inputs and inverse probabilistic inputs), as we consider more and more methods.
The chart at the bottom represents the absolute call frequency for each method.
In each chart, the data series corresponding to the \emph{sample runs} is depicted in \textbf{\color{blue} blue}, the series corresponding to the \emph{probabilistic runs} in \textbf{\color{OliveGreen} green}, and the series for the \emph{inverse probabilistic runs} in \textbf{\color{orange} orange}.

\subsubsection{Research Question 1}

In order to argue for a positive answer for \textbf{(RQ1)}, we need to compare the statistical distributions resulting from our strategies.  Such a comparison is a notoriously difficult problem, with the common advice being to run a visual test~\cite{buja2009statistical}.  We need to be able to see a pattern in frequency calls such that the accumulated curves for the \emph{sample} runs and the \emph{probabilistic} runs roughly match.\footnote{Note that in every chart the curve for the {\color{OliveGreen} probabilistic runs} is always smooth. This is a result of the sorting being done on the frequency of calls on this variant.}  It can be seen that this match does hold in all JSON examples very closely, and to a further extent also by the Rhino JS interpreter and CSSValidator.

We also perform a statistical analysis on the distributions to increase the confidence in our conclusion.
To this end we aim at performing a distribution fitness test (KS - Kolmogorov-Smirnov) on the sample vs. the probabilistic call distribution; and on the sample vs. the inverse probabilistic distribution.
It must be noted that the KS test aims at determining whether the distributions are \emph{exactly} the same, whereas we want to ascertain if they are \emph{similar} or \emph{dissimilar}.
KS tests are \emph{very sensitive} to small variations in data, which makes it, in principle, inadequate for this objective.
In this work, we employ the approach used in~\cite{fan1994testing}---we first estimate the kernel density functions of the data distributions, which smoothens the estimated distribution.
Then, we bootstrap and resample new data on the kernel density estimates, and perform the KS test on the bootstrapped data.
Results are shown in \Cref{tbl:distrcomparison}, column (C). Results range from strong (in blue) to inconclusive (in orange and red) on relating the sample to the probabilistic data\footnote{In the case of the Jackson exemplars, frequencies for the sample calls are all close to zero, which makes the data inadequate for the KS test.}.

\begin{result}
In almost subjects, the ``more of the same'' method call frequency distribution matches the distribution in the sample.
In the other cases, results are mostly inconclusive.
\end{result}

In the cases of most discrepancy, they can be explained looking at the absolute frequencies, looking for the spikes that cause the curves to mismatch.
In the case of the Rhino JS interpreter, two spikes distinguish themselves clearly, with frequencies hovering around $24\%$ and $39\%$ of all calls---that is, they account for more than \emph{half} of the total method calls on the sample.
In looking at the data, it turns out that these spikes correspond to methods {\small \<org.mozilla.javascript.TokenStream.getChar>} \\and {\small \<{org.mozilla.javascript.TokenStream.<init>}>}.
This is explained by the fact that the samples have real-world properties, such as sensibly-named (and therefore longer) variables and methods, whereas our approach tends to generate much shorter names.
In the case of the CSSValidator subject, the situation is similar.
The frequency chart shows three distinct spikes which correspond to methods {\small \<util.Utf8Properties.continueLine>} (8\%), {\small \<util.U\\tf8Properties.loadConversion>} (19\%) and {\small \<util.Ut\\f8Properties.removeWhiteSpaces>} (23\%), which deal with utf-8 conversions.
Again, the load on those methods is larger, as names are longer in real-world samples.

\def\freqfigure#1#2{
\begin{figure}[pt]
	\centering
	\includegraphics*[width=0.9\linewidth,trim=0 25px 0 0]{#1_ACCUM.png}
	\includegraphics*[width=0.9\linewidth,trim=0 25px 0 0]{#1_RAWFREQ.png}
	\caption{Call frequency analysis for #1}
	\label{#2}
        \vspace{-\baselineskip}
\end{figure}
}

\freqfigure{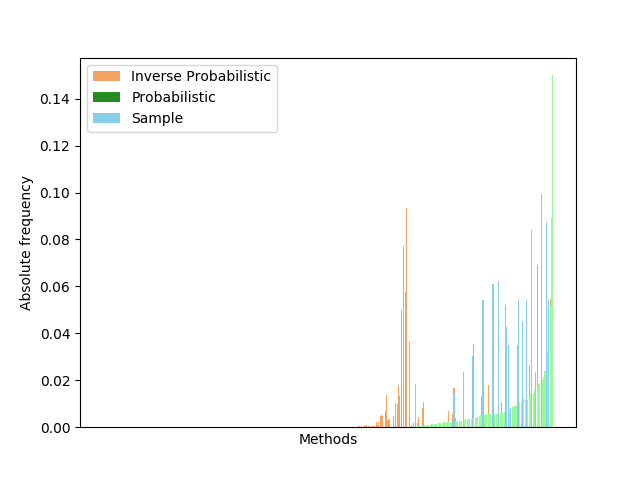}{fig:json_results_gson}
\freqfigure{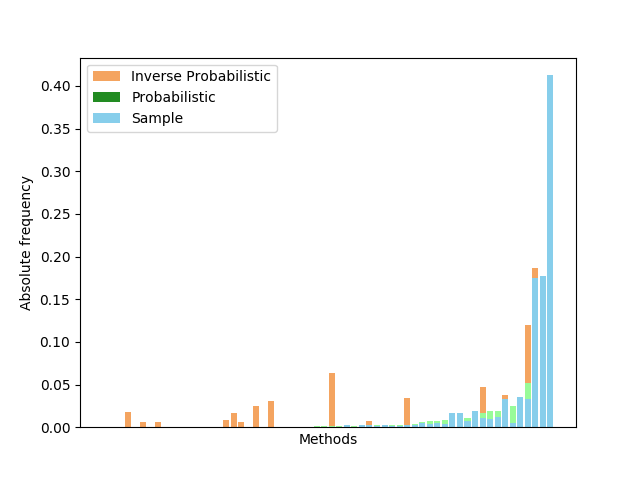}{fig:json_results_jsonjava}

\freqfigure{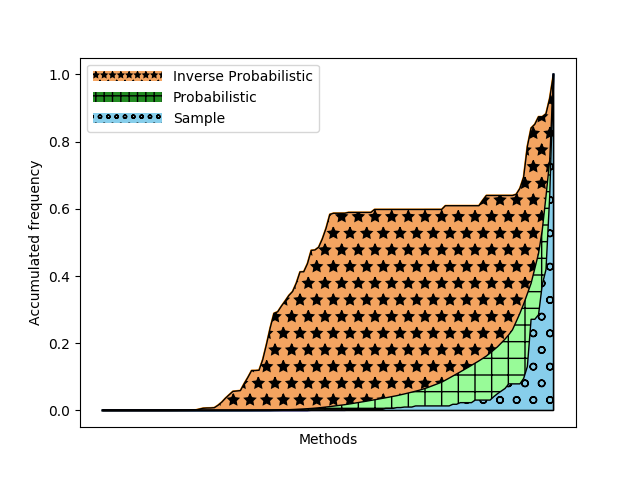}{fig:json_results_minimaljson}

\freqfigure{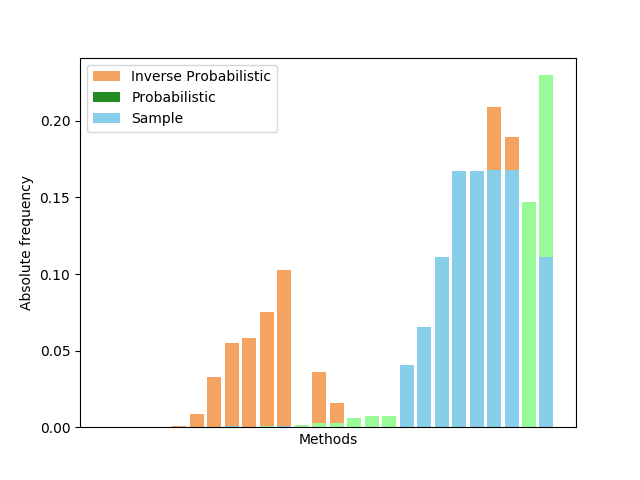}{fig:json_results_jsonsimple}
\freqfigure{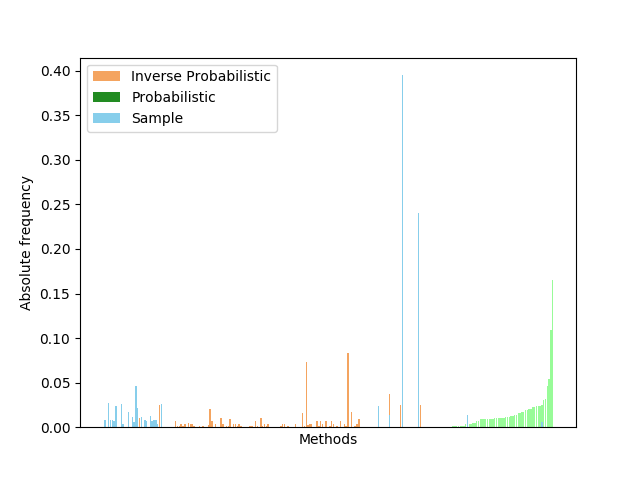}{fig:js_results_rhino}
\freqfigure{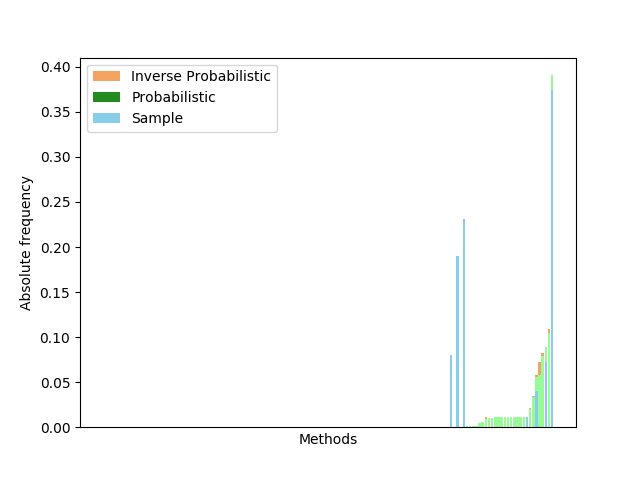}{fig:css_results_cssvalidator}

\begin{table}[ht]
	\scriptsize%
	\centering
	\begin{tabular}{@{}lrrrr@{}}
		{\bf Subject} & \thead{A} & \thead{B} & \thead{C} & \thead{D} \\
		\hline
		Argo & 10 & +3.32\% & {\color{blue} (0.37, \num{1.33e-6})} & (0.57, \num{4.41e-15}) \\
		Genson & 24 & +8.16\% &  {\color{blue}(0.13, 0.34)} & (0.51, \num{3.70e-12}) \\
		Gson & 21 & +6.14\% & {\color{blue}(0.08, 0.89)} & (0.43, \num{9.12e-9}) \\
		JSONJava & 6 & +10.00\% & {\color{blue}(0.15, 0.19)} & (0.61, \num{3.28e-17}) \\
		Jackson & 20 & \textbf{+2.22\%} & N/A & N/A \\
		JsonToJava & 11 & +12.09\% & {\color{blue}(0.23, \num{8.21e-2})} & (0.59, \num{3.96e-16}) \\
		MinimalJson & 25 & +20.49\% & {\color{orange}(0.36, \num{2.85e-6})} & (0.58, \num{1.34e-15}) \\
		Pojo & 18 & +8.78\% & {\color{blue}(0.21, 0,02)} & (0.53, \num{4.26e-13}) \\
		json-simple & 8 & \textbf{+30.77\%} & {\color{orange}(0.27, \num{1.03e-2})} & (0.58, \num{1.34e-15}) \\
		Rhino & 13  & +5.22\% & {\color{red}(0.70, 1.57e-22)}  & (0.51, \num{3.69e-12}) \\
		cssValidator & 17 & +11.88\% & {\color{orange}(0.49, \num{2.95e-11})} & (0.48, \num{8.08e-11}) \\
	\end{tabular}
	\caption{(A): number of methods called by the inverse approach that are \emph{never} called by the samples - (B): percentage increase of (A) w.r.t. sample - (C, D): smoothed bootstrapped Kolmogorov-Smirnov tests for distributions (C): sample vs. probabilistic; (D): sample vs. inverse (test statistic, p-value).}
	\label{tbl:distrcomparison}
\end{table}

\subsubsection{Research Question 2}

In this case, we want to check if we see a \emph{markedly different} accumulated frequency between the samples and the inputs generated by the inverse probabilistic approach.
Again, it can easily be seen that in almost all charts this is the case, except for the CSSValidator subject.

\begin{result}
With the exception of CSSValidator, the method call frequency distribution of ``less of the same'' is markedly different from the distribution in the sample.
\end{result}

Most intriguing for this subject is the fact that the curves for the probabilistic and inverse-probabilistic generations fit each other almost perfectly.
An in-depth analysis of the learned probabilistic grammar, however, revealed that the probabilistic grammar is \emph{almost uniform}, which explains why the inverted probabilistic grammar would look very much alike.
The results for \textbf{(RQ1)} and \textbf{(RQ2)} on this subject suggest that the approach does not work very well when the samples induce an almost-uniform grammar; and that, apparently, the original CSS grammar and real world samples are such that they don't allow for much variety, therefore resulting in such an almost-uniform grammar.

\begin{result}
A ``less of the same'' strategy works best if the element distributions in the sample input is non-uniform.
\end{result}

Further evidence on the power of the ``less of the same'' approach is shown in \Cref{tbl:distrcomparison}.
Column (A) shows the absolute number of methods that were frequently called in the ``less of the same'' inputs that were not called at all in the samples.
Column (B) shows this data as a percentage of the methods not covered in the sample.
In column (D) we perform the distribution fitting test between the sample call distribution and the inverse probabilistic one.
All results show the distributions are markedly different with strong statistical evidence.

\subsection{Threats to Validity}
\label{subsec:threats}
\paragraph*{Internal validity}
The main threat to internal validity is the correctness of our implementation.
Namely, whether our implementation does indeed learn a probabilistic grammar corresponding to the distribution of the real world samples used as training set.
Unfortunately, this problem is not a simple one to resolve.
The probabilistic grammar can be seen as a Markov chain, and the aforementioned problem is equivalent to verifying that its equilibrium distribution corresponds to the posterior distribution of the real world samples.
The problem is two-fold: first, the number of samples necessary in order to ascertain the posterior distribution is inordinate.
Second, even if we had a chance to process such a number of inputs, or if the posterior distribution were otherwise known, it might well be the case that the probabilistic grammar actually has no equilibrium distribution\footnote{A Markov chain with multiple bottom (isolated) connected components will have no equilibrium distributions. It can be easily the case that a grammar has such a multiplicity of connected components.}.
However, our tests on smaller and simpler grammars suggest that this is not an issue.

A second internal validity threat is present in the technique we use for controlling the size of the generated samples.
As described before, a sample's size is defined in terms of the number of expansions in its parsing tree.
In order to control the size, we keep track of the number of expansions generated.
Once this number crosses a certain threshold (if it actually crosses it at all), all open derivations are closed via their shortest path.
This does introduce a bias in the generation that does not exactly correspond to the distribution described by the probabilistic grammar.
The effects of such a bias are difficult to determine, and merit further and deeper study.
However, not performing this termination procedure would render useless any approach based on probabilistic grammars.

\paragraph*{External validity}
Threats to external validity relate to the generalizability of the experimental results.
In our case, this is specifically related to the subjects used in the experiments.
We acknowledge that we have only experimented with a limited number of input grammars.
However, we have selected the subjects with the intention to test our approach on practically relevant input grammars with different complexities, from small to medium size grammars like JSON; and rather complex grammars like JavaScript and CSS.
As a result, we are confident that our approach will also work on inputs that can be characterized by context-free grammars with a wide range of complexity.
However, we do have evidence that the approach does not seem to be generalizable to combinations of grammars and samples such that they induce the learning of an almost-uniform probabilistic grammar.

\paragraph*{Construct validity}
The main threat to construct validity is the metric we use to evaluate the similarity between test suites, namely method call frequency.
While the uses of coverage metrics as adequacy criteria is extensively discussed by the community \cite{anand2013BCCCGHHMOE,bertolino2007,zhu1997HM}, their binary nature (that is, we can either report \emph{covered} or \emph{not covered}) makes them too shallow to differentiate for behavior.
The variance intrinsic to the probabilistic generation makes it very likely that at least one sample will cover parts of the code unrelated to those covered by the rest of the suite.
Indeed, we carried out coverage-based experiments on our probabilistically and inverse-probabilistically generated suites, and this metric turned out to be inadequate, as we did not find significant differences when looking at binary notions of coverage.

\section{Related Work}
\label{sec:related-work}

{\bf Software Test Generation.}  The aim of \emph{software test generation} is to find a sample of inputs that induce executions that sufficiently cover the possible behaviors of the program---including undesired behavior.
Modern software test generation relies, as surveyed by Anand et al.~\cite{anand2013BCCCGHHMOE} on \emph{symbolic code analysis} to solve the path conditions leading to uncovered code~\cite{visser2004,BohmePNR17, cadar2008,cadar2008GPDE,KhurshidPV03, li2013SWL,christakis20160W,tillmann2008H}, \emph{search-based approaches} to systematically evolve a population of inputs towards the desired goal~\cite{mcminn2011, fraser2011, panichella2017,MaoHJ16}, random inputs to programs and functions~\cite{pacheco2007, miller1990} or a combination of these techniques~\cite{godefroid2005, sen2005,boehme2017PNR,rawat2017JKCGB,toffola2017SP}.
Additionally, machine learning techniques can also be applied to create test sequences~\cite{LiuZPZMZ17,SuMCWYYPLS17}.
All these approaches have in common that they do not require an additional model or annotations to constrain the set of generated inputs; this makes them very versatile, but brings the risk of producing false alarms---failing executions that cannot be obtained through legal inputs.

{\bf Grammar-Based Test Generation.}  The usage of grammars as \emph{producers} was introduced in 1970 by Hanford in his \emph{syntax machine}~\cite{hanford1970}.
Such producers are mainly used for testing compilers and interpreters: CSmith~\cite{yang2011csmith} produces syntactically correct C~programs, and LANGFUZZ~\cite{holler2012fuzzing} uses a JavaScript grammar to parse, recombine, and mutate existing inputs while maintaining most of the syntactic validity.
GENA~\cite{guo2013Q,guo2015Q} uses standard symbolic grammars to produce test cases and only applies stochastic annotation during the derivation process to distribute the test cases and to limit recursions and derivation depth.
Grammar-based white-box fuzzing~\cite{godefroid2008} combines grammar-based fuzzing with symbolic testing and is now available as a service from Microsoft.
As these techniques operate with system inputs, any failure reported is a true failure---there are no false alarms.
None of the above approaches use probabilistic grammars, though.

{\bf Probabilistic Grammars.}  The foundations of probabilistic grammars date back to the earliest works of Chomsky~\cite{chomsky1957}.
The concept has seen several interactions and generalizations with physics and statistics; we recommend the very nice article by Geman and Johnson~\cite{geman2000} as an introduction.
Probabilistic grammars are frequently used to analyze ambiguous data sequences---in computational linguistics~\cite{manning1999} to analyze natural language, and in biochemistry~\cite{sakakibara1994} to model and parse macromolecules such as DNA, RNA, or protein sequences.
Probabilistic grammars  have been used also to model and produce input data for specific domains, such as 3D~scenes~\cite{liu2014} or processor instructions~\cite{cekan2017}.

The usage of probabilistic grammars for test generation seems rather straightforward, but is still uncommon.
The \emph{Geno} test generator for .NET programs by L\"ammel and Schulte~\cite{laemmel2006} allowed users to specify probabilities for individual production rules.
This approach, in contrast to the one we present in this paper, does not use existing samples to learn or estimate probabilities.
The test case generation~\cite{kifetew2014TT,kifetew2017TT} and failure reproduction~\cite{kifetew2014JTOT} approaches by Kifetew et al. combine probabilistic grammars with a search-based testing approach.
The results~\cite{kifetew2017TT} show that the combination produces a large percentage of correct inputs and, based on the fitness function, produces a high-branch coverage.
However, due to the search-based nature of the approach, a large number of system evaluations to determine the fitness of the generated test cases are required.
The approach by Poulding et al.~\cite{feldt2013P,poulding2015ACH} uses a stochastic context-free grammar for statistical testing.
The goal of this work is thus to correctly imitate the operational profile and consequently the generated test cases are similar to what one would expect during normal operation of the system.

{\bf Mining Probabilities.}  Related to our work are approaches that mine grammar rules and probabilities from existing samples.
Patra and Pradel~\cite{patra2016} use a given parser to mine probabilities for subsequent fuzz testing and to reduce tree-based inputs for debugging~\cite{herfert17PP}.
Their aim, however, is not to produce inputs that would be similar or dissimilar to existing inputs, but rather to produce inputs that have a higher likelihood to be syntactically correct.
This aim is also shared by two \emph{grammar mining} approaches: GLADE~\cite{bastani2017synthesizing} and Learn\&Fuzz~\cite{godefroid2017learnandfuzz}, which
learn producers from large sets of input samples even without a given grammar.

All these approaches, however, share the problem of producing only ``more of the same''---they can only focus on common features rather than uncommon features, creating a general ``tension between conflicting learning and fuzzing goals''~\cite{godefroid2017learnandfuzz}.
In contrast, our work can specifically focus on uncommon inputs---that is, the complement of what has been learned.

Like us, the Skyfire approach~\cite{WangCWL17} aims at also leveraging uncommon inputs for probabilistic fuzzing.
Their idea is to learn a probabilistic distribution from a set of samples and use this distribution to generate seeds for a standard fuzzing tool, namely AFL~\cite{AFLFuzz}.
Here, favoring low probability rules is one of many heuristics applied besides low frequency, low complexity, or production limits.
The tool requires, however, the specification of a context-dependent grammar.
Although the tool has shown good results for XML-like languages, results for other, general grammar formats such as JavaScript are marked as ``preliminary'' only, though.

{\bf Mining Grammars.}  Our approach requires a grammar that can be used both for parsing and producing inputs.
While engineering such a grammar may well pay off in terms of better testing, it is still a significant investment in the case of specific domain inputs where such a grammar might not be immediately available.
Mining input structures~\cite{lin08Z}, as exemplified using the above GLADE~\cite{bastani2017synthesizing} and Learn\&Fuzz~\cite{godefroid2017learnandfuzz} approaches, may assist in this task.
The AUTOGRAM approach by H\"oschele and Zeller~\cite{hoschele2016autogram} mines human-readable input grammars exploiting structure and identifiers of a program processing the input, which makes it particularly promising.

\section{Conclusions and Future Work}
\label{sec:conclusion}

In this paper we have presented an approach that allows engineers, using a grammar and a set of input samples, to generate instances that are either similar or dissimilar to these samples.
Similar samples are useful, for instance, when learning from failure-inducing inputs; while dissimilar samples could be used to leverage the testing approach to explore previously uncovered code.
Our approach provides a simple, general, and cost-effective means to generate test cases that can then be targeted to the commonly used portions of the software, or to the rarely used features.

Despite their usefulness for test case generation, grammars---including probabilistic grammars---still have a lot of potential to explore in research, and a lot of ground to cover in practice.
Our future work will focus on the following topics:

{\bf Deep models.}  At this point, our approach captures probabilistic distributions only at the level of individual rules.
However, probabilistic distributions could also capture the occurrence of elements in particular \emph{contexts}, and differentiate between them.
For instance, if a \term{+} symbol rarely occurs within parentheses, yet frequently outside of them, this difference would, depending on how the grammar is structured, not be caught by our approach.
The domain of computational linguistics~\cite{manning1999} has introduced a number of models that take context into account.
In our future work, we shall experiment with deeper context models, and determining their effect on capturing common and uncommon input features.

{\bf Grammar learning.}  The big cost of our approach is the necessity of a formal grammar for both parsing and producing---a cost that can boil down to 1--2 programmer days if a formal grammar is already part of the system (say, as an input file for parser generators), but also extend to weeks if it is not.
In the future, we will be experimenting with approaches that \emph{mine grammars} from input samples and programs~\cite{bastani2017synthesizing, godefroid2017learnandfuzz, hoschele2016autogram, lin08Z} with the goal of extending the resulting grammars with probabilities for probabilistic fuzzing.

{\bf Debugging.}  Mined probabilistic grammars could be used to characterize the features of failure-inducing inputs, separating them from those of passing inputs.
Statistical fault localization techniques~\cite{wong2016survey}, for instance, could then identify input elements most likely associated with a failure.
Generating ``more of the same''  inputs, as in this paper, and testing whether they cause failures, could further strengthen correlations between input patterns and failures, as well as narrow down the circumstances under which the failure occurs.

We are committed to making our research accessible for replication and extension.
The source code of our parsers and production tools, the raw input samples, as well as all raw obtained data and processed charts is available as a replication package:

\begin{center}
        \url{https://tinyurl.com/inputs-from-hell}
\end{center}

\balance
\printbibliography
\end{document}